\PassOptionsToPackage{pdftex, xetex}{graphicx}
\PassOptionsToPackage{usenames,dvipsnames,svgnames,table}{xcolor}
\documentclass[letterpaper, 10pt, conference]{IEEEtran} 
\IEEEoverridecommandlockouts
\usepackage{pratik}

\title{
\LARGE \tbf{Scalable Reinforcement Learning Policies for Multi-Agent Control}
}

\author{Christopher D. Hsu, Heejin Jeong, George J. Pappas, and Pratik Chaudhari
\thanks{Department of
Electrical \&  Systems Engineering and the GRASP
Laboratory at the University of Pennsylvania.
\href{mailto:chsu8@seas.upenn.edu}{chsu8@seas.upenn.edu},
\href{mailto:heejinj@alumni.upenn.edu}{heejinj@alumni.upenn.edu},
\href{mailto:pappasg@seas.upenn.edu}{pappasg@seas.upenn.edu},
\href{mailto:pratikac@seas.upenn.edu}{pratikac@seas.upenn.edu}}
\thanks{The interested reader can find the implementation at \url{https://github.com/christopher-hsu/scalableMARL}}
}

\begin{document}

\maketitle

\begin{abstract}
We develop a Multi-Agent Reinforcement Learning (MARL)
method to learn scalable control policies for target tracking.
Our method can handle an arbitrary number of pursuers and targets;
we show results for tasks consisting up to 1000 pursuers tracking 1000 targets.
We use a decentralized, partially-observable Markov Decision Process framework to
model pursuers as agents receiving partial observations (range and bearing) about
targets which move using fixed, unknown policies.
An attention mechanism is used to parameterize the value function of the agents; this
mechanism allows us to handle an arbitrary number of targets.
Entropy-regularized off-policy RL methods are used to train a stochastic policy,
and we discuss how it enables a hedging behavior between pursuers that leads to
a weak form of cooperation in spite of completely decentralized control execution.
We further develop a masking heuristic that allows training on smaller problems with
few pursuers-targets and execution on much larger problems.
Thorough simulation experiments, ablation studies, and comparisons to state of the art
algorithms are performed to study the scalability of the approach and robustness
of performance to varying numbers of agents and targets.
\end{abstract}

\section{Introduction}
This paper studies a multi-agent control problem where 
a team of pursuers tracks multiple targets. This problem
has a wide variety of
applications that range from defense~\cite{Wise2006, Hilal2013AnIS},
search and rescue~\cite{Kumar2004}, to
environmental monitoring~\cite{Dunbabin2012}.
\begin{wrapfigure}{r}{0.5\linewidth}
    \centering
    \includegraphics[width=\linewidth]{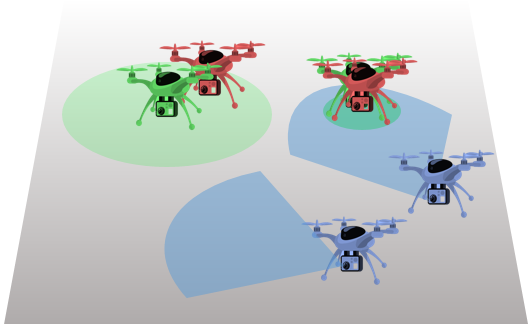}
    \caption{Multi-agent pursuer, multi-target tracking task: a team of blue agents with limited range and bearing sensing (blue cones) track red targets; the belief of
    target states, shown in green, is computed using the observation history
    of the targets.}
    \label{fig:trackingfig}
\end{wrapfigure}
Our work is motivated by the following challenges in this problem.
First,
the number of targets, or even
the number of pursuers may not be known a priori.
In particular, for learning-based approaches which the
present paper focuses on, the number of agents at deployment can be
different (typically \emph{larger}) than that at training time.
Second, control polices should ideally be decentralized to enable execution
without communication overheads and latency. However, a trade-off in this setting is that it is difficult to design controllers that can benefit from cooperative strategies to improve performance.
Our approach to resolving these challenges is based on the following key ideas.

\heading{1. Off-policy Reinforcement Learning for multi-agent control problems.}
We model pursuers as agents that observe the distance and bearing of targets within their sensing  radius; their goal is to localize target states while targets follow a fixed, but unknown policy.
We use entropy-regularized off-policy reinforcement learning (RL) to learn the pursuers' value function which aims to minimize the uncertainty of the targets' states maintained by the pursuers' belief system.
For the large-scale multi-agent problems we are interested in, continuous-control policies are challenging to learn. Therefore, we have the pursuers choose actions from a discrete set of motion primitives; our experiments indicate that discrete actions are enough to learn performant policies.

\heading{2. Building policies that can handle an arbitrary number of pursuer agents and targets.}
Each pursuer follows a control policy with the goal of maintaining a Gaussian belief distribution for each target; pursuers receive noisy measurements about target locations when targets are within its sensing radius. Given this setting, it is desirable for the RL-based policy to not be limited by a specific number of targets or pursuer agents. We achieve this by parameterizing a value function using a self-attention-based neural architecture and sharing it amongst agents. This architecture has two powerful properties: (i) it creates a permutation-invariant embedding of the input state-vector (the beliefs) which makes the value function invariant to the ordering of the targets in the pursuer's vicinity, and (ii) it can handle an input vector of an \emph{arbitrary} size, which provides the ability of deploying the trained policy on larger problems, with many more targets. Since execution is decentralized across pursuers, the approach is naturally deployable with different numbers of pursuer agents. Furthermore, our approach can seamlessly handle tasks where the number of agents or targets changes during deployment.

\heading{3. Using stochastic policies to obtain a weak form of cooperation.}
Centralized planning/execution and communication are bottlenecks for cooperation in
large-scale multi-agent problems. 
We develop an effective workaround for this issue by training
a \textit{stochastic} RL policy in centralized fashion but executing it in decentralized fashion;
neither local observations nor control actions are shared amongst pursuers during deployment.
This paradigm forces multiple pursuers to \emph{hedge} their control actions. For example, 
two pursuers have two targets in their vicinity, since the policy is trained to minimize the average
uncertainty across all targets, instead of both non-communicating pursuers heading towards the closest target, entropic
regularization in the policy encourages them to also seek the further target with a small probability.
This weak, but extremely useful, form of cooperation improves tracking performance as compared
to deterministic policies.

\heading{4. Heuristics for Scalability.}
Training large multi-agent policies is difficult:
we observe that even state of the art sample-efficient off-policy RL algorithms fail to learn performant policies in the MARL setting.
We develop a heuristic that takes a policy trained on a smaller problem, say 4 agents and 4 targets, and executes this policy on a larger problem by masking far away targets. We show that this masking is particularly effective; we can execute policies trained on the smaller problem, with minimal degradation, on problems as large as 1000 pursuers tracking 1000 targets. This heuristic allows us to address a dramatic distribution shift in the number of targets: it essentially scales down the large and complex observation space at test time into something that is close to what the pursuer policy was trained for. Experiments show that this heuristic performs much better than state of the art MARL algorithms for large problems.

\section{Problem Formulation}

Consider the multi-agent multi-target setup where the goal of a team of $N$
homogeneous pursuers is to acquire information about $M$ homogeneous targets.
The $i^{\trm{th}}$ agent with state $x^i_t \in \mathbb{R}^2 \times SO(2)$,
and the $j^{\trm{th}}$ target with state
$y^j_t \in \mathbb{R}^2 \times SO(2)$,
follow a discrete-time dynamical model.
Given the initial state for the pursuer
$x^i_0 \in \mathcal{X}$, an initial target state $y^j_0$ and a horizon $T$,
the pursuer agent chooses control actions $u^i_t$ to maximize
the mutual information between $y^j_t$ and the measurement history denoted
by $z_{1:t}$. Mutual information between the $i^{\trm{th}}$ pursuer
and the $j^{\trm{th}}$ target $I(y^j_t, z_{1:t}) = \trm{KL}\big(p(y^j_t, z_{1:t}),\ p(y^j_t) p(z_{1:t})\big)$
is the Kullback-Leibler (KL) divergence
$\trm{KL}(q, p) = \int \trm{d} q \log (q/p)$
of the joint distribution
and the product of the marginals. The objective for localizing
target states is
\begin{align}
\label{MI}
    \max_{u(\cdot)}\ 
    \sum_{i=1}^N \sum^T_{t=0} \sum_{j=1}^M I(y^j_{t}, z_{1:t}\ |\ x^i_{1:t})
\end{align}
such that $x^i_{t+1} = f(x^i_t, u^i_t)$, $y^j_{t+1} = g(y^j_t)$ and
$z_{t} = h(x^i_t, y^1_t, \ldots, y^M_t)$ and $u^i_t = u(z_{1:t})$
for $t \in \cbr{0,T}$. The functions $f(\cdot)$ and $g(\cdot)$
are the dynamical models of the pursuers and the targets respectively,
and $h(\cdot)$ is the observation model. Note that the measurement history $z_{1:t}$ is shared among the agents. Further decentralization can occur by performing distributed information filtering~\cite{Schlotfeldt2018}.

\subsection{Multi-agent Formulation}

We consider the decentralized partially observable multi-agent setting where pursuers cannot observe the full state of the environment. Each agent receives observations of targets in its vicinity. Therefore, the task is defined as a decentralized partially observable Markov Decision Process (Dec-POMDP) represented by the tuple, $(G, S, A_i,\mathcal{Z}_i, P, R, \gamma)$, where $G$ is a set of agents, $S$ is a set a states, $A_i$ is a discrete set of actions for each agent $i$, $\mathcal{Z}_i$ is a set of observations for each agent $i$, and $P, R, \gamma$ are the state transition probability kernel, the global reward, and the discount factor, respectively. In this work, we consider problems where $G$, $S$, and $\mathcal{Z}$ can be infinite to account for scalable teams in a continuous domain. Optimally solving Dec-POMDPs is a hard combinatorial problem that involves searching through all the agent's histories. This process results in an exponentially growing policy space as the time horizon increases, making the policy search quickly intractable~\cite{Dibangoye2013}. The following simplifying assumptions are made to keep the problem tractable.

\subsubsection{Partial Observability}

In the partially observable setting, the agent does not have access to the ground truth target states. As a substitute, the agent maintains a belief on target states and solves a continuous belief MDP~\cite{Smallwood1973}. We denote such a belief distribution for the $j^{th}$ target as $B(y_{j,t}) = p(y_{j,t}|z_{1:t},x_{1:t})$ and its predicted distribution for the subsequent step as $\Bar{B}(y_{j,t+1}) = p(y_{j,t+1}|z_{1:t},x_{1:t})$~\cite{jeong2019learning}. Assuming $y_{t+1}$ is independent of $x_{1:t}$, the optimization problem can be reduced to minimizing the cumulative differential entropy, $H(y_{t+1}|z_{1:t}, x_{1:t})$~\cite{atanasov2013information}. When the belief is Gaussian, $B(y_{j,t}) = \mathcal{N}(\hat{y}_{j,t},\Sigma_{j,t})$,
    \begin{align}
        \scalemath{0.9}{
        H(y_{t+1}|z_{1:t}, x_{1:t}) = \frac{1}{2}\log\big((2\pi e)^M \det(\Sigma_t)\big)
        }
    \end{align}
    the optimization problem (\ref{MI}) becomes:
    \begin{align} \label{DE}
        \min_{\pi} \sum_{t = 0}^T \log\det(\Sigma_t)
    \end{align}
	
At each time step $t$, the agent at state $x_t$, chooses a control input $u_t$ from the policy $\pi$, based on the predicted belief of the target, $\Bar{B}(y_{t+1})$. In order to maximize the objective (\ref{MI}), the agent receives a measurement $z_{t+1}$ from the sensor. At the same time, the true target state transitions from $y_t \xrightarrow{} y_{t+1}$ and if it is observed by the agent, the corresponding belief distribution is updated with the new measurement. This process is repeated until the time horizon $T$. In the case of multiple agents or multiple targets, at each time step, all agents and targets will transition before the environment transitions from $t \xrightarrow{} t+1$.

\subsubsection{Decentralization}

Consider a team of $N$ agents pursuing $M$ targets where each agent can make control actions independently of its team. A tractable solution to this problem is through the use of the centralized training, decentralized execution parameter sharing approach~\cite{Gupta2017}. Assuming that the agents are homogeneous, during centralized training, agents share the parameters of a single value network and an experience replay buffer. This formulation reduces the policy space that the algorithm has to search through because it optimizes a single, shared policy rather than multiple individual policies~\cite{Tan93multi-agentreinforcement}. This approach also mitigates the non-stationarity issue in multi-agent training by exposing the policy to all of the agents' experiences simultaneously~\cite{terry2020revisiting}. A shared policy also alleviates computational issues in multi-agent centralized critic algorithms that suffer from the curse of dimensionality problem as the number of agents increases~\cite{lowe2020multiagent}. A parameter-shared policy can be executed in decentralized fashion at test time for \emph{any} number of agents.

\subsection{Off-policy Reinforcement Learning}
In reinforcement learning (RL), the goal is to learn the action-value function which is defined as the expected reward obtained using the policy $\pi$, from an initial state $x$, with control action $a$. Off-policy RL is a popular technique to estimate the action-value function. It minimizes the expectation of the 1-step temporal difference (TD) error. Off-policy methods are named so because they maintain a dataset $\mathcal{D} = \{(s_t, a_t, r_t, s_{t+1})\}_{t=0,\dots}$ collected using a (behavior) policy and estimate the value function of the current policy $\pi$ using this data. We will parametrize the value function using a deep network with parameters $\theta$. The optimal parameters $\theta^*$ can be found by performing stochastic gradient descent to minimize the Huber loss applied to the TD error. In the sequel, we design a simulation environment for multi-agent problems which allows us to study complex multi-agent interactions across diverse scenarios in the target tracking problem. This simulator is used to train our off-policy RL methods. We next discuss RL-specific details of this simulation setup.

\subsubsection{Observations}

In the partially observable target-tracking task, exact target states are not known to the agents. Each agent maintains a belief over target states~\cite{jeong2019learning, jeong2020learning} that is updated using a Kalman filter as new observations about the target are received. The dynamics of targets is unknown to the tracking agent, and more importantly, target states are independent of the agent's control input. The value function and the control policy of the $i^{\text{th}}$ agent therefore take the predicted belief of target $j$ at the \emph{next} time instant $t+1$ as the input at time $t$. This is neat way to ensure that the RL policy has access to the target tracking information maintained in the Kalman filter without having to relearn it from scratch, which would require a large amount of data. We also transform the belief of the $j^{\text{th}}$ target in agent $i$'s local frame ($\hat{y}^{(x_{i})}$), this ensures that there is no distribution shift between observations collected by different agents. The state is
\[
    \scalemath{1}{
    s_j := \big[\hat{y}^{(x_{i})}_{j,r}, \hat{y}^{(x_{i})}_{j,\theta}, \dot{\hat{y}}^{(x_{i})}_{j,r}, \dot{\hat{y}}^{(x_{i})}_{j,\theta}, \log\det \Sigma_{j}, \mathbb{I}(y_{j}\in \mathcal{O}(x_t)) \big].
    }
\]
The quantities $\hat{y}^{(x_{i})}_{j,r}$ and $\hat{y}^{(x_{i})}_{j,\theta}$ denote polar coordinates of the mean of target $j$'s belief. We also use their derivatives as a part of the state. Target belief covariance, $\Sigma_{j}$, is represented with the differential entropy formulation (\ref{DE}). We model agents with a limited sensing range using the Boolean function $\mathbb{I}(\cdot)$: it returns $1$ if the true target state is in the vicinity of the agent, and zero otherwise. The combined state denoted by $s_t$ of $M$ targets is the concatenation of the states of individual targets, and it is a vector of $6M$ entries.

We are motivated by the realization that if one were to use a multi-layer perceptron (MLP) for parameterizing the value function, the number of targets $M$ would have to be fixed between training and testing. Further, and this is much more debilitating, the order in which states of individual targets are concatenated together would also have to be fixed between training and testing. Ideally, we would rather train on smaller problems with few agents where it is easy to gather rich data from the simulation and execute the same policy on problems with more agents at test time. We therefore think of the observations as a set of $M$ elements and design an embedding of this set as the state for the RL policy, (i) that is permutation-invariant, i.e., it does not depend upon the order in which observations are concatenated together, and (ii) that can handle an arbitrary number of targets. Section~\ref{Attention} discusses how a self-attention-based architecture is used to achieve these desiderata.

\subsubsection{Rewards}
We would like to find the optimal policy $\pi^*$ that maximizes the cumulative mutual information (\ref{MI}) between the $i^{\trm{th}}$ pursuer and the $j^{\trm{th}}$ target, this amounts to minimizing the objective (\ref{DE}). We approximate the objective as a discounted sum by defining the reward as the average uncertainty over all the targets such that $R(s_t,a_t) = -\frac{1}{M}\sum_{j=0}^M \log\det(\Sigma_{t+1}^j)$, resulting in a value function:
\begin{equation}
\label{reward}
    \scalemath{0.9}{
    V^{\pi}(s) = -\mathbb{E}_{\pi}\Big[\frac{1}{M}\sum_{t=0}^{T-1} \sum_{j=0}^M \gamma^t\log\det(\Sigma_{t+1}^j)\ |\ s_0=s \Big]
    }
\end{equation}
Observe that the reward given to each agent depends upon the performance of all the other agents, this is desirable because the value function and the policy are shared between agents and also because doing so makes the policy $\pi^*$ responsive to how the other agents undertaking the same policy are tracking their respective targets.

\subsubsection{Discrete Action Space}
Continuous-control RL policies are challenging to learn in multi-agent settings. Our goal is to understand the multi-agent aspects of the problem and in order to so, we choose a discrete action space using motion primitives: $A = \big\{(v,\omega)|v\in \{0,0.67,1.33,2.0\}[m/s], \omega\in \{-\pi/4, 0, \pi/4\}[\trm{rad}/s]\big\}$. This set of actions is rich enough to lead to performant tracking policies, and yet, it is small enough for us to be able to learn RL policies for large, challenging problems.

\section{Approach}

\subsection{Self-Attention-based Model Architecture} \label{Attention}
    We require a policy that can be trained on smaller problems and executed at test time on larger problems with more targets in order to improve sample efficiency and alleviate debilitating restrictions such as requiring specific input ordering~\cite{Mern2020}. Therefore, for this setting, we think of the observations as a set of elements that is fed into an embedding (i) that is permutation-invariant, i.e., the embedding is invariant to changes in the ordering of observations, and (ii) that can handle an arbitrary number of elements in the observation set. In this work, the policy is built with an encoder-decoder style model architecture as described by DeepSets~\cite{zaheer2017deep}, filled with self-attention layers~\cite{vaswani2017attention}. Self-attention layers provide the desired property of permutation-invariance by highlighting the actionable information in the observation~\cite{Soatto2009}. The second property of handling an arbitrary number of elements is taken care of by the summation in the DeepSets architecture, i.e., any additional elements are simply part of the linear combination of the embedding space.

    
    \subsubsection{Permutation input representations and embeddings of sets}
    DeepSets~\cite{zaheer2017deep} show that a function $\phi(A)$ operating on a set A is invariant to permutation of the elements in A iff it can be decomposed as $\phi(A)\equiv \rho(\sum_{a\in A}\psi(a))$ for some functions $\psi$ and $\rho$. Both $\psi$ and $\rho$ can be learnt to build invariance and the summation enables $\phi(A)$ to handle sets with varying number of elements.
    
    \subsubsection{Attention}
    The self-attention mechanism~\cite{vaswani2017attention} is a powerful way for the value function to learn to pay attention to parts of the input that are more relevant to the output. Through the structure of learning embeddings, attention amounts to mapping a query, $Q$, and a set of key, $K$, value, $V$ pairs to an output. Structured as a linear combination of $V$, attention is able to learn these embeddings by computing the dot product $QK^T$ which measures how compatible they are. The attention module $\omega(d_{q}^{-\frac{1}{2}}QK^T)V$ gives more weight to the key that has a large dot product with the query vector, i.e., the output is computed as a weighted sum of the values, a measure of the compatibility between the query and the corresponding key. $\omega$ is an activation function, e.g. softmax, and $\frac{1}{\sqrt{d_q}}$ is a scaling factor. Finally, the self-attention module can be improved by considering higher order interactions of $Q$ and $K$ by projecting the inner product across multiple sub-spaces, creating the Multi-Head Attention Block (MAB).

\subsection{Maximum entropy policies} \label{max-entropy}
    In order to learn a stochastic policy, the maximum entropy-regularized objective~\cite{ZiebartMaxEntropy}  looks to balance maximizing the expected return and entropy, i.e, to be successful at the task while acting as randomly as possible. The entropy of the policy is controlled by a temperature parameter, $\alpha$, that reflects the idea that high entropy equates to high temperature and high randomness and vice versa. The entropy H of a distribution P over a random variable x is given as:
    \begin{align}
        H(P) = \mathbb{E}_{x\sim P}[-\log P(x)]
    \end{align}
    At each time step, the agent gets a bonus reward proportional to the entropy of the policy. The RL problem can be reformulated with a maximum entropy objective:
    \begin{align}
        J(\pi) = \sum_{t=0}^{T-1}\mathbb{E}_{(s_t,a_t)\sim \pi}[r(s_t,a_t) - \alpha \log\pi(a_t|s_t)]
    \end{align}
    
    This objective favors stochastic policies by augmenting the equation with an expected entropy over the policy. Stochastic policies have conceptual and practical advantages. First, the policy is incentivized to explore more widely, while giving up on unpromising trajectories. Secondly, the policy can capture multiple modes of optimal behavior, i.e., when there are equally attractive actions, the policy places an equal probability mass to each of those actions. Lastly, these stochastic policies have shown to substantially improve exploration and are therefore beneficial for cooperative tasks.

\subsection{Details of the off-policy training implementation} \label{training}

    We utilize and augment the state-of-the-art off-policy method Clipped Double Q-Learning~\cite{fujimoto2018addressing} to provide the structure in which we learn the deterministic and stochastic policies. The original Double Q-Learning algorithm learns two independent estimates of the true Q value parameterized by neural networks denoted as $Q_{\phi_1}$ and $Q_{\phi_2}$. This framework is able to reduce overestimation of the Q value by taking the minimum value of the two independent estimates. The author suggests that this method provides higher values to states that have lower variance estimation error, leading to more stable learning. The double Q-learning update is highlighted below.
    \begin{align}
        \scalemath{0.9}{
        Q^*(s,a) \xleftarrow{} r + \gamma\min_{k=1,2}Q_{\phi_k'}(s',\arg\max_{a'} \sum_{k=1,2} Q_{\phi_k}(s',a'))
        }
    \end{align}
    Double Q-learning learns a deterministic policy due to the action being chosen by taking the $\arg\max(Q(s,a))$. Standard deterministic policies utilize an $\epsilon$-greedy exploration schedule during training. 
    
    To learn a stochastic policy,
    we augment the standard double Q-learning formulation with an entropy-regularized objective~\cite{haarnoja2018soft}. This algorithm (see Alg. \ref{PS-sdql})
    is denoted as Soft Double Q-learning and balances the expected return and entropy during exploration. The soft double Q-learning update is
    \begin{align} \label{softupdate}
        \scalemath{0.9}{
         Q^*(s,a) \xleftarrow{} r + \gamma\min_{k=1,2} \E_{a' \sim \pi} \big[Q_{\phi_k'}(s', a')
        - \alpha \log \pi(a'|s')\big]
        }
    \end{align}
    With a stochastic policy, actions are chosen by sampling from a multinomial defined by the log softmax of the Q function values.
    During exploration, a larger entropy is used to encourage the policy to sample actions that might not necessarily have the largest value. This helps the policy learn about less frequented parts of the state space thereby improving exploration. Other implementations of maximum entropy objectives have used fixed or learned schedules to control $\alpha$ in order to promote exploration vs. exploitation. 

\subsection{Algorithm}
        We adjust the standard single-agent learning algorithm, Soft Double Q-learning, with the parameter sharing approach~\cite{Gupta2017} to provide the benefits of centralized training with decentralized execution for multi-agent learning. In Alg. \ref{PS-sdql}, the addition of parameter sharing to Soft Double Q-learning can be seen in the for loop starting in line \ref{lst:parametersharing}, where we are sampling the environment with each agent. For each environment step, we feed each of the agent's local observations through the shared networks to produce a unique experience to be stored. Only when all the agent's have taken their action with the current policy, does the full environment transition from $t \to t+1$. The second half of the algorithm, starting line \ref{lst:update}, updates the value functions with stochastic gradient descent. We sample from a replay buffer that includes the experiences from all the agents. This procedure helps to nullify the non-stationarity problem of decentrally trained systems. When training is complete, each of the agents receives a copy of the learned policy that can be used to execute decentrally.


    In order to improve generalization, $n$ number of agents in the current team and $m$ number of targets are sampled uniformly randomly during training to create different numbers of cooperative agents in the team as seen in line \ref{lst:sample}. This has two benefits. First, it shows flexibility of the algorithm as $n$ and $m$ fluctuate during training. Second, it helps sample diverse training episodes and
    expands the task space, resulting in a more generalizable and robust policy.

\begin{algorithm} [!htpb] 
    \footnotesize
    \setstretch{0.9}
    \caption{Parameter-Shared Soft Double Q-Learning} \label{PS-sdql}
    \begin{algorithmic}[1]
        \State Initialize value networks $Q_{\phi_1}, Q_{\phi_2}$
        \State Initialize target networks $\phi_1' \xleftarrow{} \phi_1, \phi_2' \xleftarrow{} \phi_2$
        \State Initialize N and M, max number of agents and targets
        \State Initialize replay buffer $\mathbb{G}$
        \For{each iteration}
            \State Randomly sample $n \in [1,N]$ and $m \in [1,M]$  \label{lst:sample}
            \For{each environment step}
            \For{$i \xleftarrow{} 0,1,\dots n$} \label{lst:parametersharing}
                \State Observe state $s_t$ and sample $a_t \sim \pi(a_t|s_t)$
                \State Execute $a_i$; observe next state $s_i'$, reward $r_t$
                \State Store $(s_{i},a_{i},r_t,s_{i}')$ in replay buffer $\mathbb{G}$
            \EndFor
            \EndFor
            \For{each update step} \label{lst:update}
                \State Sample mini-batch of N:
                \State $\quad g_t = (s,a,r,s') \sim \mathbb{G}$
                \State Compute soft target Q value using \ref{softupdate}
                \State Perform clipped gradient descent step on
                \State $\quad \sum_{k=1,2}\text{huber}(Q_{\phi_k}(s,a) - Q_{\text{targ}}(s,a))$
                \State Update target network parameters:
                \State $\quad \phi_k' \xleftarrow{} \tau\phi_k + (1-\tau)\phi_k'$
            \EndFor
        \EndFor
    \end{algorithmic}
    \end{algorithm}

\subsection{Heuristic Mask}

In this paper, the heuristic mask is used during evaluation to improve scalability of trained policies. At every time step, each agent's observation state is masked to expose the $k$ nearest target beliefs based on its range measurement, $\hat{y}_{j,r}$.

\begin{algorithm} [H]
    \footnotesize
    \setstretch{0.9}
\caption{Heuristic Mask for Scalability} \label{heuristic}
\begin{algorithmic} 
\For{each environment step}
\For{$i \xleftarrow{} 0,1,\dots, n$}
    \State $s_t \xleftarrow{} \text{mask}_{k, \hat{y}_{j,r}}\{s_t(\hat{y}_{1,r},\ldots,\hat{y}_{M,r})\}$
\EndFor
\EndFor
\end{algorithmic}
\end{algorithm}

\section{Experimental validation}

\subsection{Setup}

    The multi-agent multi-target tracking task results in a highly stochastic environment. We randomly initialize agent, target, and belief locations. Targets move about randomly with their fixed double integrator with Gaussian noise model. The belief model and the observation model also have noise components, compounding the randomness of the environment. With this in mind, to lower training variance and focus on tracking rather than a search task, we define a restricted random initialization for target locations and their corresponding beliefs. All the agents' positions, $x_{0,\dots,N}$, are randomly initialized within the given map, avoiding other teammates. The targets' positional component of $y_{0,\dots,M}$ are randomly initialized 5-10 meters away from an uniformly randomly sampled agent. The targets' initial velocity components are set to 0.0. The Gaussian belief of the targets' locations are initialized 0-5 meters away from their respective target with covariance $\Sigma$. Both agents and targets have a maximum velocity of 2 m/s. The time horizon $T$ is 200 steps. Finally, each agent has an observation range of $10m$ and bearing of $\frac{\pi}{4}$ for a total of $25\pi m^2$ of coverage. See~\cite{jeong2020learning} for more details of the target tracking environment, including agent, observation, and target models.
\subsubsection{Framework}
\begin{wrapfigure}{r}{0.6\linewidth}
        \centering
        \includegraphics[width=\linewidth]{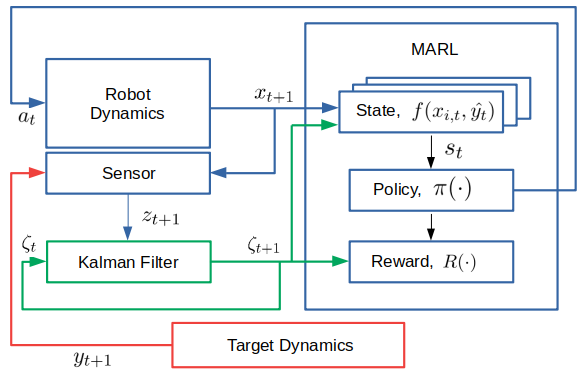}
        \caption{\textbf{Multi-Agent Target Tracking Framework.} The MARL block is built of any number of homogeneous agents with the same dynamics. Each robot receives a local observation state of target beliefs from the centrally stored Kalman Filters. The policy, $\pi$, can easily be exchanged with models such as the DeepSets self-attention, deterministic and stochastic models or the MLP model.}
        \label{fig:marlframework}
    \end{wrapfigure}

    The framework shown in Fig. \ref{fig:marlframework}, expands upon a single agent strategy~\cite{jeong2019learning} with the use of parameter sharing for multiple agents.

\subsubsection{Evaluation methodology}

    In multi-agent multi-target domains, the balance between the number of agents and the number of targets is critical to the behaviors. We will train on the tasks defined by $n \in [1,N]$ agents and $m \in [1,M]$ targets and evaluate, without retraining, on other task spaces with or without a heuristic mask. To denote the difference between policies, for example, a deterministic self-attention-based policy trained on the task space of $[1,4]$ agents and $[1,4]$ targets will be "det-4a4t" and a stochastic self-attention-based policy trained on the same task space will be "stoch-4a4t". When evaluating policies on larger task spaces, we also scale the map to fit the density of the originally trained 4a4t task where the empty map has an area of $2500 m^2$. With each agent able to observe an area of $25\pi m^2$, we maintain the same density when scaling to 8, 20, 100, and 1000 agents. It is important in RL to train the policies on multiple seeds in order to reduce the possible variance caused by random initialization~\cite{Mania2018}. We train each policy on multiple random seeds and evaluate on 50 episodes per task for 5 seeds.
    
\subsubsection{Assumptions}
    
    In the current framework, we designate a centrally stored Kalman filter that each of the agents has access to and can update. We assume that the data association problem is solved, i.e. if an agent observes a target, the corresponding belief of that target is updated correctly. When a different agent queries the central Kalman filter, it receives the updated information from the rest of the agents. Although this part of the system is centralized, local observations and control actions are not shared amongst the team. Therefore, this formulation allows for an easy extension to a distributed Kalman filtering paradigm~\cite{Schlotfeldt2018} for further decentralization. 
    
    Targets move around with a double integrator model with Gaussian noise and an additional noise term when faced with obstacles, i.e. the walls of the map.  The Kalman filter is updated with the double integrator, however, agents are unaware of this extra noise term. Therefore, when targets bounce off the walls, belief updates become considerably inaccurate leading to a more challenging task.

\subsection{Baseline methods and experiments}
\begin{figure} [!htpb] 
        \centering
        \includegraphics[scale=0.3]{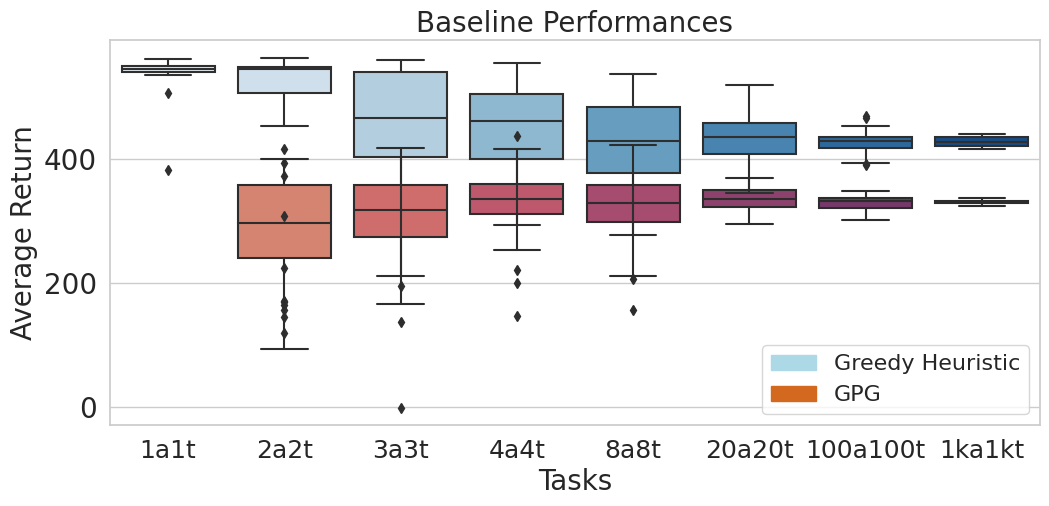}
        \caption{Task labels, "1a1t" and "1ka1kt", denote the task of the environment where one agent tracks one target and 1000 agents track 1000 targets, respectively. In blue, the trained stochastic-4a4t policy, during evaluation, is masked with the greedy heuristic of $k=1$, where $k$ is the number of targets to consider. It can be seen that the greedy heuristic baseline does a very good job of extrapolating to tasks outside of the original training space without much performance degradation by reducing the task to a single-agent single target tracking environment. In red, a state-of-the-art graph policy gradient (GPG) algorithm~\cite{khan2019graph} seemingly scales, but suffers in performance. (evaluation on 50 episodes/task for 5 seeds). 
        }
        \label{fig:baseline}
    \end{figure}

    \emph{Greedy heuristic.}
    Our first baseline is a greedy heuristic mask that reduces the multi-agent multi-target tracking problem
    into a single-agent single-target problem; each agent's observation state is masked to only have the nearest target's belief.
    This is a simple baseline and comparisons against it are a sanity check.  With the greedy heuristic, note that based on the random initial locations of the targets, two agents can greedily track the same target, leaving targets free. This phenomenon can be seen in the variance of the middle sized tasks in Fig.~\ref{fig:baseline}. The greedy heuristic opens up the possibility for performance improvement. 
    \emph{However, we observe
    in Fig.~\ref{fig:baseline} that even this baseline is better than state-of-the-art graph-based policy gradient (GPG~\cite{khan2019graph})}.
    Note that these authors employ an idea that has a similar setting: training a graph CNN on a smaller task
    and evaluating the policy on larger tasks without retraining.

    \emph{Performance of the graph policy gradient (GPG) algorithm.} In Fig.~\ref{fig:baseline}, GPG is not evaluated on the 1a1t task as with 1 agent there is no graph to define. Although GPG seemingly scales, it consistently falls into a local optima even when trained on multiple seeds and training lengths. We surmise that GPG, designed for the fully observable setting, cannot handle policy learning in a partially observable setting where training is less stable.

\begin{figure} [!htpb]
        \centering
        \includegraphics[scale=0.25]{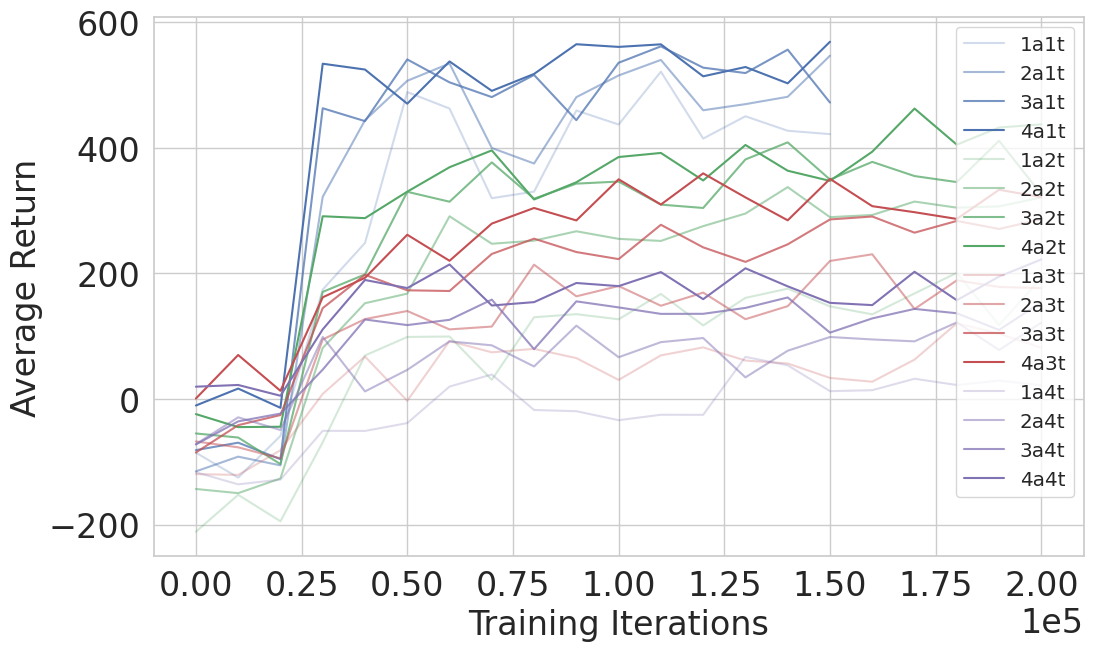}
        \caption{Online returns of 4 different MLP-based policies based on the number of targets, averaged across 5 seeds. The curve associated with "1a1t" is the cumulative sum of returns collected on an single-agent single-target task. Teams are more successful with more agents (color shades) and suffer when tracking more targets (color-coded).}
        \label{fig:ma_online}
    \end{figure}

    
    \emph{Multi-layer perceptron (MLP).} We train MLP policies with parameter sharing between pursuer agents using double Q-learning. A different MLP policy is needed depending on the number of targets (1-4), however, parameter sharing allows for a varying number of agents. Baseline performance is shown in Fig.~\ref{fig:ma_online}. As expected, when the number of targets increase, the expected average return received decreases. 
    Intuitively, this result stems from the fact that even if the uncertainty of 3 of 4 targets is low, a high uncertainty of the last target strongly affects the overall averaged reward.
    
\subsection{Main results}
\subsubsection{Deterministic vs Stochastic Policies}

In real-time, requiring the knowledge of the number of targets at initialization is a severe limitation, therefore, we present deterministic and stochastic self-attention-based policies. Both policies are trained with parameter sharing while the former uses double Q-learning and the latter uses soft double Q-learning. This architecture has the property to allow for changes in number of agents or targets during evaluation. We train a deterministic policy and a stochastic policy that each generalizes over the 16 tasks of $n \in [1,4]$ agents and $m \in [1,4]$ targets in the multi-agent multi-target tracking environment. 

\begin{figure} [!htpb]
        \centering
        \includegraphics[scale=0.3]{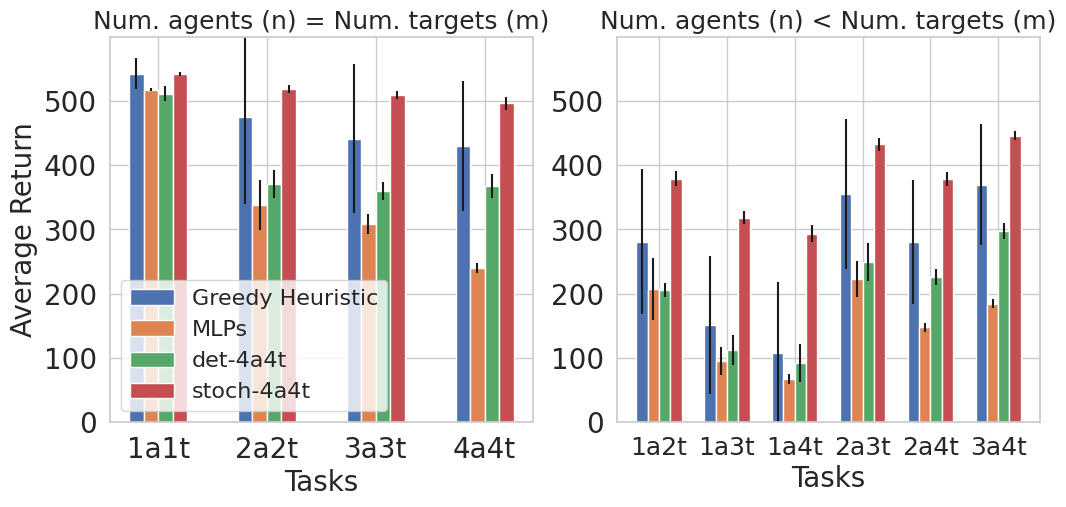}
        \caption{Comparing the architecturally different policies when $n \leq m$. The stochastic policy (stoch-4a4t) outperforms all other policies with minimal variance. (evaluation on 50 episodes/task for 5 seeds).}
        \label{fig:comparingnm}
    \end{figure}

\begin{figure} [!htpb]
        \centering
        \includegraphics[width=0.8\linewidth]{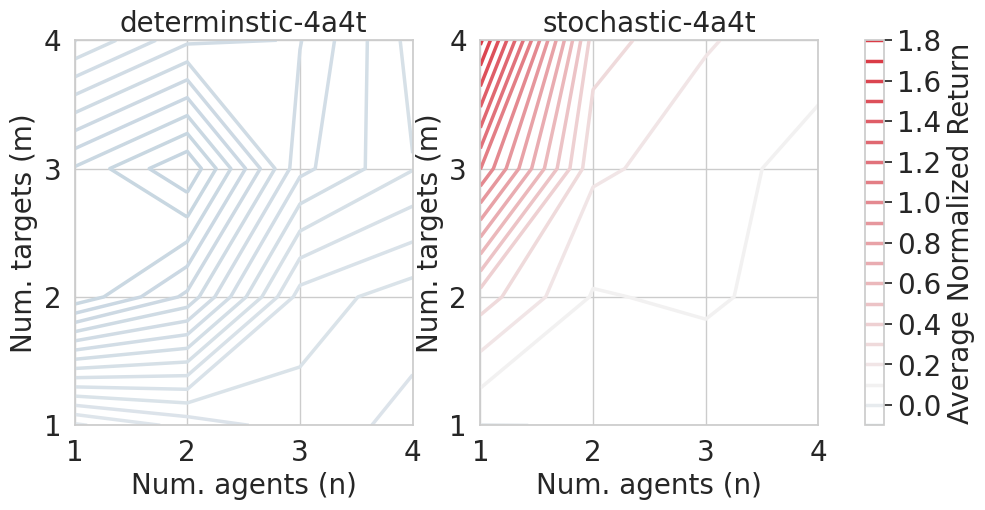}
        \caption{Contour plots of the deterministic and stochastic policy trained and evaluated on $n\in[1,4]$ agents tracking $m\in[1,4]$ targets. The returns are normalized against the greedy heuristic baseline and averaged across the seeds evaluated on 50 episodes/task. Lines on the red spectrum denote returns greater than the baseline (white, 0.0) while blue is less than. Stochastic policies stand out against the baseline in the most difficult parts of the task space.}
        \label{fig:contour}
    \end{figure}

\begin{figure} [!htpb]
        \centering
        \includegraphics[scale=0.23]{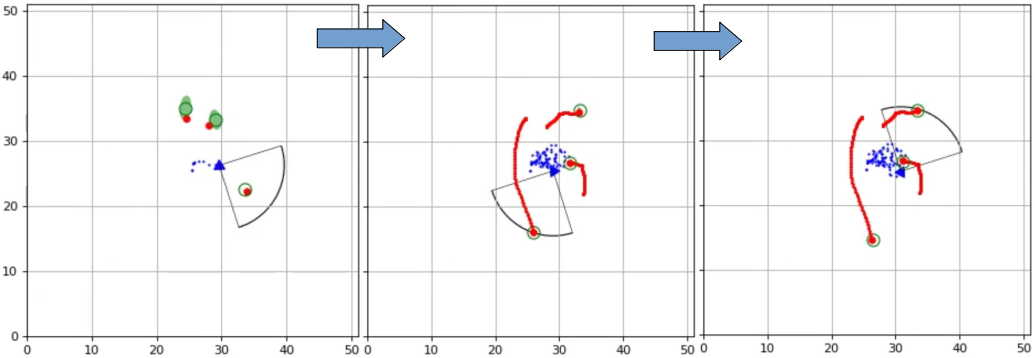}
        \caption{Example evaluation episode of the stochastic policy maintaining low uncertainty of the 3 targets.}
        \label{fig:stoch1a3t}
    \end{figure}

    From Fig.~\ref{fig:comparingnm} and Fig.~\ref{fig:contour}, we surmise that the self-attention-based policy that takes deterministic actions, results in a policy that can be extremely greedy and sub-optimal; each agent acting individually acquiring short term rewards rather than as a cooperative team gaining long term high rewards. Although it outperforms the 4 MLP-based policies, it under performs against the greedy heuristic baseline which has an intrinsic sense of target assignment.
    Therefore, we also developed a stochastic self-attention-based policy
    which outperforms the greedy heuristic across all tasks with much lower variance.     
    This phenomenon occurs due to the policy acting more intelligently to minimize uncertainty of all targets and forgoes short term rewards for long term higher team rewards, whereas the greedy heuristic policy focuses on the closest target (highly dependent on initializations) and allows the uncertainty of the rest of the targets to explode. 
    We observe a weak form of cooperation from the stochastic policy as it is able to hedge it control actions while robustly tracking across different tasks.
    Stochastic policies separate itself from the rest of the policies in the more difficult task space of $n < m$ and in simpler parts of the task space, it performs the optimal policy. 
    Episodic visualization can be seen in Fig. \ref{fig:stoch1a3t}.

    In Fig. \ref{fig:stoch1a3t}, we illustrate the stochastic policy in a 1 agent tracking 3 targets scenario. The single agent acts intelligently, rotating its sensing region amongst the targets to maintain a strong belief. The greedy heuristic policy (not shown) would focus on the closest target in question and maintain constant observation as it only considers the one-on-one scenario, leaving the other two targets to gain large uncertainties.

\subsection{Ablation experiments}

\subsubsection{Scalability}

\begin{figure}[!htpb]
        \centering
        \includegraphics[width=0.9\linewidth]{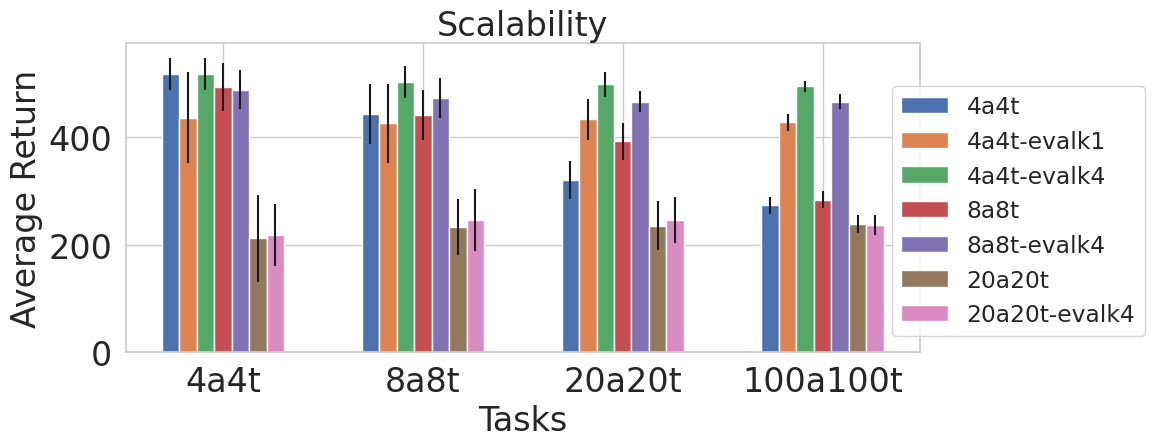}
        \caption{Comparing scalability of policies trained on larger task spaces. Heuristically masked policies (-evalk) outperform their unmasked policies. Policies evaluated with masks of $k=4$ outperform the mask of $k=1$. Training on smaller task spaces leads to more stable training and higher average returns. Heuristic masking further improves scalability. (evaluation on 50 episodes/task for 5 seeds).}
        \label{fig:scalability}
    \end{figure}

    
    We study in Fig. \ref{fig:scalability} whether it would be better to retrain policies on the larger task space or
    simply allow our stochastic policy to extrapolate with either the full observation state or a heuristic mask. We compare a stochastic policy trained
    on the task space of $n \in [1, 4]$ and $m \in [1, 4]$ denoted as ``4a4t'',
    a stochastic policy trained on $n \in [1, 8]$ and $m \in [1, 8]$ denoted as ``8a8t'',
    and a stochastic policy trained on $n \in [1, 20]$ and $m \in [1, 20]$ denoted as ``20a20t''. 
    These trained policies are then evaluated, without retraining, on the larger task spaces with either the full observation state or a heuristically masked state, e.g. "-evalk1" (greedy heuristic, $k=1$) or "-evalk4" (a mask of $k=4$). 
    
    As we saw previously, there are pros and cons of using the greedy heuristic (4a4t-evalk1): the former gives agents a specific objective while the latter precludes the opportunity for better performance compared to one-on-one tracking. Fig.~\ref{fig:scalability} shows
    that applying a $k=4$ heuristic on the stochastic 4a4t and 8a8t policies outperforms all other policies as they balance limiting the agent's objectives and utilizing the stochasticity of the policy as seen in Fig. \ref{fig:comparingnm}.
    When scaling to larger spaces, a heuristic mask handles the distribution shift enacted by the large number of targets by allowing the policy to return back to the distribution space that it was originally trained within, a space with high performance.
    
    Evaluating policies with the full observation state results in a degradation in performance as the task space expands past their training space. Policies 4a4t and 8a8t marginally outperform the greedy heuristic baseline for tasks within the training space but degrade outside of it.
    Meanwhile, the 20a20t policy does not work well even on tasks inside
    its training space. This result demonstrates the challenges with training MARL systems; 
    the task space here is 25 times larger that of 4a4t (400 tasks vs. 16 tasks).
    The larger task space policies were trained to convergence, however, the local-optimum that they settle in is not optimal.
    We conclude that training on larger task spaces in MARL is difficult, and simply
    using a heuristic on a well-trained 4a4t policy provides better performance than directly training large MARL systems.

\subsubsection{Cooperative behavior due to the stochastic policy}

\begin{figure}[!htpb]
        \centering
        \includegraphics[width=0.8\linewidth]{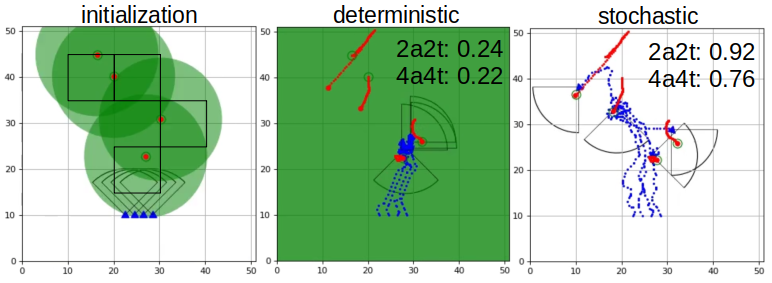}
        \caption{Deterministic and stochastic policies were evaluated on tasks where the targets were randomly initialized somewhere in the corresponding black box. Out of 50 episodes, how many times are the agents able to observe all the targets in the map? Results of the corresponding policy and task are shown on the graph. Not shown, the greedy policy in both tasks received a score of 0.0.}
        \label{fig:cooperation}
    \end{figure}

    In Fig. \ref{fig:cooperation}, we look to display the cooperative-like behaviors that emerge from stochastic policies and compare it with the deterministic policy. We set up 2a2t and 4a4t scenarios where the team of agents is initialized towards the bottom of the map and they must search to observe all the targets. The targets are initialized in $10 \times 10 m^2$ boxes with increasingly further ranges to force the agents to cooperate.
    
    The stochastic policies learn something more significant than just tracking the closest target. Not only are able to find the further targets, they are able to self delegate which target is assigned to whom, confirming our intuition that they have some sense of cooperation. Unsuccessful episodes occurred when the stochastic policy agents take too long. The target has already moved too far away from the belief for the agent to observe it and the agent is forced to randomly search.

\subsubsection{GRU vs KF}

\begin{figure}[!htpb]
        \centering
        \includegraphics[width=0.75\linewidth]{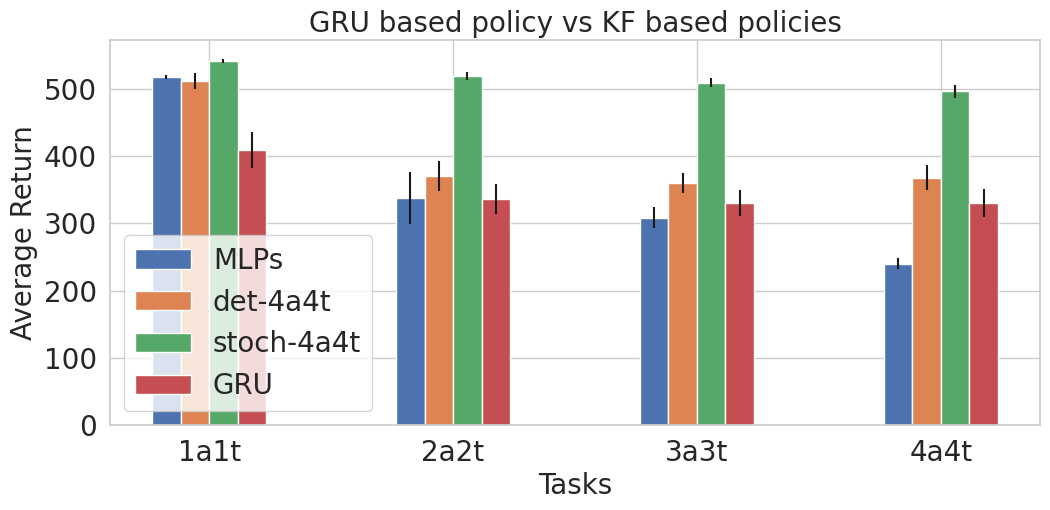}
        \caption{Comparison of policies utilizing a KF vs a GRU; the GRU under performs the policies utilizing a KF and self-attention network. (evaluation on 50 episodes/task for 5 seeds).}
        \label{fig:gru}
    \end{figure}

    So why did we choose to use a Kalman filter (KF) to update target beliefs? Why not make the policy a fully neural network based model? We evaluate a policy, shown in Fig. \ref{fig:gru}, that uses a Gated Recurrent Unit (GRU) to maintain the belief statistics of the target. The output of the GRU to the stochastic policy is the mean and covariance of the target beliefs. The policy should learn to use those statistics in order to predict the locations of the targets and therefore perform just as well as previously shown policies.
    
    The lack of performance can be hypothesized as the following. In comparison with the GRU, the KF formulation maintains a statistic for each target. The agent can simply query the associated KF for the target beliefs. On the other hand, the single GRU must maintain statistics for all the targets. If desired, it is possible to have a GRU for each target in order to mimic the KF formulation. However, during training, as the number of targets increase, there is an ever growing parameter space to have to learn, reducing the power of the scalability in our formulation. With a KF, there is nothing to learn and with more targets, it is simply a matter of storing the additional belief statistics.

\section{Related Work}

    Multi-agent control for pursuit-evasion has seen a wide range of perspectives,
    see for instance sampling-based algorithms for few agents which allow for
    real-time implementation~\cite{Karaman2010}. Analytical solutions for general dynamical
    systems are difficult to compute, except for special cases, e.g., the homicidal chauffeur~\cite{isaacs1954}.
    In pursuit-evasion problems, perimeter defense and multi-agent settings have been handled by decomposing the task into two-player games and reducing the defense strategy to an assignment problem~\cite{Chen2014}. More explicit cooperation can be performed by decomposing into smaller local-games played by a subset of agents~\cite{Shishika2019}.
    
    Information acquisition methods that build control policies to minimize the entropy in the
    estimation task are close to our methods;~\cite{atanasov2013information, Atanasov2015}. These works
    develop non-myopic control policies by minimizing the uncertainty in the target state, conditioned on past measurements.
    Decentralized policies can be obtained in MARL using independent Q-functions~\cite{Tan93multi-agentreinforcement}, but training these policies is challenging due
    to non-stationarity. Centralized policies alleviate non-stationarity, but have to pay the cost
    of exponentially large multi-agent action-spaces~\cite{lowe2020multiagent, foerster2017counterfactual}. Works like QMIX~\cite{rashid2018qmix}, and QTRAN~\cite{son2019qtran}, formulate a centralized value representation to approximate the collection of individual action-values obtained by each agent during training.
    Wider literature using graphs networks~\cite{khan2019graph, li2021deep} and permutation invariant structures~\cite{liu2019pic} provides similar claims of scalability and cooperation. However, these works evaluate their algorithms in fully observable settings where training is stable. We observe that when~\cite{khan2019graph} is executed with partial observations, it is difficult to learn a performant policy.
    %
    There are also directions that focus on emergent behaviors in multi-agent systems to build cooperative and competitive strategies~\cite{tampuu2015multiagent}, identify emergent roles~\cite{wang2020roma}, coordination and emergent use of tools~\cite{baker2020emergent}.
    
    Similar to us,~\cite{iqbal2019actorattentioncritic} utilized an attention network in an MARL framework to get permutation invariance. However, they used attention in a centralized critic which forces the input space to scale linearly with the number of agents whereas ours is independent of that dimension.  
    \cite{carion2019structured} uses a structured prediction approach to assign agents to tasks. The assignment policy is well learned on small problem instances and evaluated on problems with more agents; they show results for about 80 agents. Other strategies for task allocation in swarm intelligence with multi-agent and multi-target tracking have been also investigated~\cite{Low04,Low05autonomicmobile}.
    Another strategy uses a curriculum learning paradigm to scale MARL~\cite{long2020evolutionary}. Through an evolutionary approach, they are able to introduce more agents into the systems and select the policies that best adapt; they show results with 24 agents.

\section{Conclusions}

We presented an off-policy reinforcement learning method for multi-agent target tracking and demonstrated that our decentralized policy is able to scale to an arbitrary number of agents, e.g., up to 1000 pursuers tracking 1000 targets. The key innovation of this approach it to train on a lower dimensional task, e.g., with fewer agents, and execute---without retraining---on larger tasks. Architectural innovations in the policy such as self-attention layers and a masking heuristic at test time enable this and demonstrate how off-policy RL methods can be used for large-scale multi-agent problems.

The central challenge in multi-agent problems lies in the fact that agents need to take decentralized control decisions to work within the constraints of communication latency and yet accomplish some form of cooperation or information sharing to be able to track multiple targets. This paper explores a heuristic to strike such a tradeoff between communication and cooperation. It trains a stochastic policy, as opposed to a deterministic policy, that helps agents hedge their control actions at test time when multiple targets are in the vicinity. This is a very weak form of cooperation, but it comes at zero communication cost. This heuristic is not well-suited for more challenging scenarios which fundamentally require cooperative tracking using multiple agents, e.g., targets whose uncertainty increases quickly. Such environments are good avenues for further exploration.


\begin{thebibliography}{10}
\providecommand{\url}[1]{#1}
\csname url@rmstyle\endcsname
\providecommand{\newblock}{\relax}
\providecommand{\bibinfo}[2]{#2}
\providecommand\BIBentrySTDinterwordspacing{\spaceskip=0pt\relax}
\providecommand\BIBentryALTinterwordstretchfactor{4}
\providecommand\BIBentryALTinterwordspacing{\spaceskip=\fontdimen2\font plus
\BIBentryALTinterwordstretchfactor\fontdimen3\font minus
  \fontdimen4\font\relax}
\providecommand\BIBforeignlanguage[2]{{%
\expandafter\ifx\csname l@#1\endcsname\relax
\typeout{** WARNING: IEEEtran.bst: No hyphenation pattern has been}%
\typeout{** loaded for the language `#1'. Using the pattern for}%
\typeout{** the default language instead.}%
\else
\language=\csname l@#1\endcsname
\fi
#2}}

\bibitem{Wise2006}
\BIBentryALTinterwordspacing
R.~Wise and R.~Rysdyk, \emph{UAV Coordination for Autonomous Target Tracking}.
  [Online]. Available: \url{https://arc.aiaa.org/doi/abs/10.2514/6.2006-6453}
\BIBentrySTDinterwordspacing

\bibitem{Hilal2013AnIS}
A.~R. Hilal, ``An intelligent sensor management framework for pervasive
  surveillance,'' 2013.

\bibitem{Kumar2004}
\BIBentryALTinterwordspacing
V.~Kumar, D.~Rus, and S.~Singh, ``Robot and sensor networks for first
  responders,'' \emph{IEEE Pervasive Computing}, vol.~3, no.~4, p. 24–33,
  Oct. 2004. [Online]. Available: \url{https://doi.org/10.1109/MPRV.2004.17}
\BIBentrySTDinterwordspacing

\bibitem{Dunbabin2012}
\BIBentryALTinterwordspacing
M.~Dunbabin and L.~Marques, ``Robots for environmental monitoring: Significant
  advancements and applications,'' \emph{IEEE Robotics and Automation
  Magazine}, vol.~19, no.~1, pp. 24--39, 2012. [Online]. Available:
  \url{https://eprints.qut.edu.au/63029/}
\BIBentrySTDinterwordspacing

\bibitem{Schlotfeldt2018}
B.~{Schlotfeldt}, D.~{Thakur}, N.~{Atanasov}, V.~{Kumar}, and G.~J. {Pappas},
  ``Anytime planning for decentralized multirobot active information
  gathering,'' \emph{IEEE Robotics and Automation Letters}, vol.~3, no.~2, pp.
  1025--1032, 2018.

\bibitem{Dibangoye2013}
J.~Dibangoye, C.~Amato, O.~Buffet, and F.~Charpillet, ``Optimally solving
  dec-pomdps as continuous-state mdps,'' vol.~55, 08 2013.

\bibitem{Smallwood1973}
\BIBentryALTinterwordspacing
R.~D. Smallwood and E.~J. Sondik, ``The optimal control of partially observable
  markov processes over a finite horizon,'' \emph{Operations Research},
  vol.~21, no.~5, pp. 1071--1088, 1973. [Online]. Available:
  \url{https://doi.org/10.1287/opre.21.5.1071}
\BIBentrySTDinterwordspacing

\bibitem{jeong2019learning}
H.~Jeong, B.~Schlotfeldt, H.~Hassani, M.~Morari, D.~D. Lee, and G.~J. Pappas,
  ``Learning q-network for active information acquisition,'' 2019.

\bibitem{atanasov2013information}
N.~Atanasov, J.~L. Ny, K.~Daniilidis, and G.~J. Pappas, ``Information
  acquisition with sensing robots: Algorithms and error bounds,'' 2013.

\bibitem{Gupta2017}
J.~K. Gupta, M.~Egorov, and M.~Kochenderfer, ``Cooperative multi-agent control
  using deep reinforcement learning,'' in \emph{Autonomous Agents and
  Multiagent Systems}, G.~Sukthankar and J.~A. Rodriguez-Aguilar, Eds.\hskip
  1em plus 0.5em minus 0.4em\relax Cham: Springer International Publishing,
  2017, pp. 66--83.

\bibitem{Tan93multi-agentreinforcement}
M.~Tan, ``Multi-agent reinforcement learning: Independent vs. cooperative
  agents,'' in \emph{In Proceedings of the Tenth International Conference on
  Machine Learning}.\hskip 1em plus 0.5em minus 0.4em\relax Morgan Kaufmann,
  1993, pp. 330--337.

\bibitem{terry2020revisiting}
J.~K. Terry, N.~Grammel, A.~Hari, L.~Santos, and B.~Black, ``Revisiting
  parameter sharing in multi-agent deep reinforcement learning,'' 2020.

\bibitem{lowe2020multiagent}
R.~Lowe, Y.~Wu, A.~Tamar, J.~Harb, P.~Abbeel, and I.~Mordatch, ``Multi-agent
  actor-critic for mixed cooperative-competitive environments,'' 2020.

\bibitem{jeong2020learning}
H.~Jeong, H.~Hassani, M.~Morari, D.~D. Lee, and G.~J. Pappas, ``Learning to
  track dynamic targets in partially known environments,'' 2020.

\bibitem{Mern2020}
J.~{Mern}, D.~{Sadigh}, and M.~J. {Kochenderfer}, ``Exchangeable input
  representations for reinforcement learning,'' in \emph{2020 American Control
  Conference (ACC)}, 2020, pp. 3971--3976.

\bibitem{zaheer2017deep}
M.~Zaheer, S.~Kottur, S.~Ravanbakhsh, B.~Poczos, R.~Salakhutdinov, and
  A.~Smola, ``Deep sets,'' 2017.

\bibitem{vaswani2017attention}
A.~Vaswani, N.~Shazeer, N.~Parmar, J.~Uszkoreit, L.~Jones, A.~N. Gomez,
  L.~Kaiser, and I.~Polosukhin, ``Attention is all you need,'' 2017.

\bibitem{Soatto2009}
S.~{Soatto}, ``Actionable information in vision,'' in \emph{2009 IEEE 12th
  International Conference on Computer Vision}, 2009, pp. 2138--2145.

\bibitem{ZiebartMaxEntropy}
B.~D. Ziebart, A.~Maas, J.~A. Bagnell, and A.~K. Dey, ``Maximum entropy inverse
  reinforcement learning,'' 2008.

\bibitem{fujimoto2018addressing}
S.~Fujimoto, H.~van Hoof, and D.~Meger, ``Addressing function approximation
  error in actor-critic methods,'' 2018.

\bibitem{haarnoja2018soft}
T.~Haarnoja, A.~Zhou, P.~Abbeel, and S.~Levine, ``Soft actor-critic: Off-policy
  maximum entropy deep reinforcement learning with a stochastic actor,'' 2018.

\bibitem{Mania2018}
H.~Mania, A.~Guy, and B.~Recht, ``Simple random search provides a competitive
  approach to reinforcement learning,'' 03 2018.

\bibitem{khan2019graph}
A.~Khan, E.~Tolstaya, A.~Ribeiro, and V.~Kumar, ``Graph policy gradients for
  large scale robot control,'' 2019.

\bibitem{Karaman2010}
S.~Karaman and E.~Frazzoli, ``Incremental sampling-based algorithms for a class
  of pursuit-evasion games,'' vol.~68, 01 2010, pp. 71--87.

\bibitem{isaacs1954}
R.~Isaacs, \emph{Differential Games I: Introduction}.\hskip 1em plus 0.5em
  minus 0.4em\relax Santa Monica, CA: RAND Corporation, 1954.

\bibitem{Chen2014}
M.~{Chen}, Z.~{Zhou}, and C.~J. {Tomlin}, ``Multiplayer reach-avoid games via
  low dimensional solutions and maximum matching,'' in \emph{2014 American
  Control Conference}, 2014, pp. 1444--1449.

\bibitem{Shishika2019}
D.~{Shishika}, J.~{Paulos}, M.~R. {Dorothy}, M.~{Ani Hsieh}, and V.~{Kumar},
  ``Team composition for perimeter defense with patrollers and defenders,'' in
  \emph{2019 IEEE 58th Conference on Decision and Control (CDC)}, 2019, pp.
  7325--7332.

\bibitem{Atanasov2015}
N.~{Atanasov}, J.~{Le Ny}, K.~{Daniilidis}, and G.~J. {Pappas}, ``Decentralized
  active information acquisition: Theory and application to multi-robot slam,''
  in \emph{2015 IEEE International Conference on Robotics and Automation
  (ICRA)}, 2015, pp. 4775--4782.

\bibitem{foerster2017counterfactual}
J.~Foerster, G.~Farquhar, T.~Afouras, N.~Nardelli, and S.~Whiteson,
  ``Counterfactual multi-agent policy gradients,'' 2017.

\bibitem{rashid2018qmix}
T.~Rashid, M.~Samvelyan, C.~S. de~Witt, G.~Farquhar, J.~Foerster, and
  S.~Whiteson, ``Qmix: Monotonic value function factorisation for deep
  multi-agent reinforcement learning,'' 2018.

\bibitem{son2019qtran}
K.~Son, D.~Kim, W.~J. Kang, D.~E. Hostallero, and Y.~Yi, ``Qtran: Learning to
  factorize with transformation for cooperative multi-agent reinforcement
  learning,'' 2019.

\bibitem{li2021deep}
S.~Li, J.~K. Gupta, P.~Morales, R.~Allen, and M.~J. Kochenderfer, ``Deep
  implicit coordination graphs for multi-agent reinforcement learning,'' 2021.

\bibitem{liu2019pic}
I.-J. Liu, R.~A. Yeh, and A.~G. Schwing, ``Pic: Permutation invariant critic
  for multi-agent deep reinforcement learning,'' 2019.

\bibitem{tampuu2015multiagent}
A.~Tampuu, T.~Matiisen, D.~Kodelja, I.~Kuzovkin, K.~Korjus, J.~Aru, J.~Aru, and
  R.~Vicente, ``Multiagent cooperation and competition with deep reinforcement
  learning,'' 2015.

\bibitem{wang2020roma}
T.~Wang, H.~Dong, V.~Lesser, and C.~Zhang, ``Roma: Multi-agent reinforcement
  learning with emergent roles,'' 2020.

\bibitem{baker2020emergent}
B.~Baker, I.~Kanitscheider, T.~Markov, Y.~Wu, G.~Powell, B.~McGrew, and
  I.~Mordatch, ``Emergent tool use from multi-agent autocurricula,'' 2020.

\bibitem{iqbal2019actorattentioncritic}
S.~Iqbal and F.~Sha, ``Actor-attention-critic for multi-agent reinforcement
  learning,'' 2019.

\bibitem{carion2019structured}
N.~Carion, G.~Synnaeve, A.~Lazaric, and N.~Usunier, ``A structured prediction
  approach for generalization in cooperative multi-agent reinforcement
  learning,'' 2019.

\bibitem{Low04}
K.~H. Low, W.~K. Leow, and M.~Jr, ``Task allocation via self-organizing swarm
  coalitions in distributed mobile sensor network,'' \emph{Proceedings of the
  National Conference on Artificial Intelligence}, 10 2004.

\bibitem{Low05autonomicmobile}
K.~H. Low, W.~K. Leow, and M.~H. Ang, ``Autonomic mobile sensor network with
  self-coordinated task allocation and execution,'' \emph{IEEE Transactions on
  Systems, Man, and Cypernetics–Part C: Applications and Reviews}, vol.~36,
  pp. 315--327, 2005.

\bibitem{long2020evolutionary}
Q.~Long, Z.~Zhou, A.~Gupta, F.~Fang, Y.~Wu, and X.~Wang, ``Evolutionary
  population curriculum for scaling multi-agent reinforcement learning,'' 2020.

\end{thebibliography}
\end{document}